\documentclass[twocolumn,prd]{revtex4}
\usepackage{graphicx}
\usepackage{epsfig}
\begin{document}

\title{Charmonium's K2 Peak}
\author{Felipe J. Llanes-Estrada}
\affiliation{Depto. F\'{\i}sica Te\'orica I, Universidad
Complutense de Madrid, 28040 Madrid Spain.}
\altaffiliation[On leave at: ]{Theory Group, Stanford Linear
Accelerator Center, 2571 Sand Hill Rd. 94025 Menlo Park CA. }
\email{fllanes@fis.ucm.es}

\begin{abstract}
The newly reported $Y(4260)$ becomes the second most massive state 
in the charmonium family. We argue that it displaces the $\psi(4415)$ as 
the (largely) $4s$ vector  charmonium state, recall $s-d$ wave 
interference to explain the lack of a signal in $e^-e^+\to \ {\rm 
hadrons}$ 
and suggest some further study avenues
that can exclude exotic meson assignments.
The absence of a $J/\psi KK$ mode can be understood, beyond 
phase space suppression, to be a consequence of chiral symmetry.
We also provide a model calculation in this sector showing that,
although forcing the fit somewhat (which suggests a small sea quark
wavefunction component), the state can be incorporated in a 
standard scheme. 
\end{abstract}
\maketitle

\section{Reported properties of $Y(4260)$ and interpretation}

Yet another unexpected state has arisen from spectroscopy at
the  B factories. The $Y(4260)$ detected at BaBar \cite{hep-ex/0506081}
adds to the tower of $\psi$ states, but unexpectedly, below
the last known resonance in this tower. 
It is common quark model  wisdom that vector 
quarkonium states come in pairs, due to the approximately equal 
energy cost of a $d$-wave and a radial excitation. A first 
glance to the new spectrum already suggests  that this state 
should be assigned the nomenclator $\psi(4 ^3S_1)$. \\
Of course, 
$L$ is not a good quantum number and a $d$ wave will be present 
to some extent.  More important is the increased splitting and 
lower mass (200 $MeV$ below the prediction of Godfrey and Isgur
\cite{Godfrey:1985xj}, that calls for a sizeable sea quark wavefunction 
component. But this mass itself can be made consistent with
a quark model by softening the string tension at 
large distances. And again the mass is not far from the 
prediction in \cite{Estrada:2000hj} (4300).

The width of the state, reported to be about $90(20)\ MeV$ is 
also consistent with expectations from the quark model. For 
example, Barnes and Swanson \cite{Barnes:2005pb}, under the 
assumption that the $\psi(4415)$ was the $ 4 ^3S_1$, evaluated 
its width at $77\ MeV$.  A value somewhat smaller is to be 
expected at the lower mass 4260, but consistent with data. Moreover we 
agree with the quantum numbers proposed by the experimental 
collaboration based on its ISR production, also consistent with
the emision of an even number of pions and chiral symmetry 
expectations.

The discovery final state $J/\psi\pi\pi$ is also typical of
$n ^3S_1\to 1 ^3S_1$ transitions, that have been studied 
in detail for the $\psi'$ and $\psi(3770)$
\cite{Brambilla:2004wf}. Their branching fractions differ
from roughly 50\% to 0.5\% by two orders of magnitude. 
This can be understood as $d$-wave suppression in the $\psi(3770)$ and 
does not apply to the present case of the $Y(4260)$. We should be 
comparing instead with the branching fraction 
of the $3S$ state, unfortunately unknown to us. If we use the $\Upsilon$ 
system as a guidance, 
$$
\frac{\Upsilon'''\to \Upsilon\pi\pi}{\Upsilon'\to 
\Upsilon\pi\pi}\simeq 0.1-0.2
$$
indicates that we should expect a branching fraction
$
B(\psi(4S)\to J/\psi \pi\pi )\simeq 2-4\% 
$
(added phase space cannot compensate the much smaller 
wavefunction overlap). From the reported 
$$
B(Y \to J\psi \pi\pi ) \Gamma(Y\to e^-e^+)\simeq 4-7 \ eV
$$
we can thus estimate  $ \Gamma(Y\to e^-e^+)\simeq 0.2-0.35\ 
keV$. On the basis of the model calculation below we would
expect a somewhat larger lepton width. 

The absolute width to $J/\psi\pi\pi$ is conceivably 1-2 $MeV$,
a factor of 4 larger than the 2S state. This is a novel effect
that can be explained within a $c\bar{c}$ model by invoking a small 
admixture of a four quark state, or in another language, a non-vanishing
coupling to the $J/\psi f_J$ channel that is open for this decay.
This is an interesting calculation and will be the subject of future 
work.

The lack of a signal in the $J/\psi KK$ channel 
is a consequence of phase space suppression that might  be 
very enhanced due to chiral symmetry.
If the decay $Y\to J/\psi M_1M_2$ was due  exclusively to
an $SU(3)$ symmetric constant vertex, then the ratio between signals
in the $J/\psi KK$ mode and the $J/\psi \pi\pi$ mode would be
$$
\frac{\int_{m_K}^{M_Y-m_K-m_{J/\psi}}dE_1 
\int_{m_K}^{M_Y-E_1}dE_2}
{\int_{m_\pi}^{M_Y-m_\pi-m_{J/\psi}}dE_1 
\int_{m_\pi}^{M_Y-E_1}dE_2}\simeq \frac{1}{4}
$$
that could very well be observed with increased statistics. 
On the other hand, in the case of a very convergent chiral 
expansion, where the first order coupling for the outgoing
meson pair would dominate the matrix element, we would have 
an extra suppression factor 
$$
\frac{f_\pi^4 (p_{K_1}\cdot p_{K_2})^2}{f_K^4 (p_{\pi_1} \cdot 
p_{\pi_2})^2} \simeq \frac{f_\pi^4 m_K^4}{f_K^4 E_{\pi_1}^2
E_{\pi_2}^2} 
$$
lowering the ratio, after phase space integration, to about 
$0.3\%$.
This number is controlled by the high momentum phase space in 
the two pion decay channel, and is very sensitive to higher 
order terms in a chiral expansion. We urge experimental 
collaborations to quote bounds on the $J/\psi KK$ channel, as
it teaches us about the quality of chiral theory applied in an
otherwise very non perturbative domain (the doublet $\psi(4S)$ 
and $\psi(3D)$ are the highest known excitation of the QCD 
string).

The $VKK$ signal in a vector to vector transition would be 
interesting by itself because in the light sector this channel 
is closed for the $\rho(1450)$, $\rho(1700)$, $\omega(1650)$,
$\phi(1680)$, and also low lying $\psi$ states. 

The isospin of $Y(4260)$ has not been reported; as no signal is yet 
observed in the more difficult $J/\psi \pi_0\pi_0$ channel (where $I=0$ 
requires a signal half of what found in the charged 
mode). 
A signal in the $J/\psi \pi^{\pm}\pi^0$ channel automatically rules out
the charmonium assignment.
 A non zero isospin would rule out the charmonium assignment and we 
should reconsider $Y$ as a likely cryptoexotic meson.

\section{Accommodating the $Y(4260)$ in the spectrum}

To see how this new resonance fits in a standard Hamiltonian
diagonalization, we recompute the vector charmonium spectrum
in the $H_{eff}$ modeling QCD in Coulomb gauge described at 
length in \cite{Estrada:1999uh}, \cite{Estrada:2001kr},
and \cite{Estrada:2004wr}.

The Random Phase Approximation diagonalization involves four 
parameters, largely constrained from other Fock space sectors. We employ a 
current $m_c$ that is dressed by means of the BCS gap equation for 
$H_{eff}$. The Coulomb gauge kernel is replaced by a momentum-space 
Cornell-type potential 
$$
V(q)= -\frac{8\pi\sigma}{q^4}- \frac{\alpha_s}{q^2}
$$
with the Coulomb tail cut-off exponentially at a momentum of 
$5\ GeV$.
The resulting spectrum for $m_c(2 \ GeV)=1 \ GeV$, $\alpha_s=0.6$,
$\sqrt{\sigma}=350\ MeV$, that are typical for this model (with the string 
tension somewhat lower) is given in table \ref{tablita}.
\begin{table}
\caption{Coulomb gauge $H_{eff}$ RPA and updated 
experimental spectrum 
\cite{Eidelman:2004wy}
incorporating the new state.
\label{tablita}}
\begin{tabular}{|cc|cc|}
\hline
State & Calc. $M(MeV)$ & State & Calc. $M(MeV)$
\\
\hline
$J/\psi(3097)$ & 3093&$\psi(4160\pm 20)  $ & 4134 \\
$\psi(3686)  $ & 3662& $Y(4260)$ & 4347 \\
$\psi(3770)  $ & 3778& $\psi(4415\pm 10)  $ & 4421 \\
$\psi(4040)  $ & 4042&  & 4597\\
               &     &  & 4661 \\
\hline
\end{tabular}
\end{table}
The new state is somewhat too light respect to the pure 
quark-antiquark assignment, which seems to be the trend for the 
other novel states in charmonium spectroscopy above the charmed
threshold.
This is even after  readjustment 
(that, in the context of this model, means lowering the string 
tension and increasing the contribution of the Coulomb potential 
to avoid generating too light masses for the low-lying states).
The calculation includes $s-d$ wave mixing and forward-backward 
RPA propagating wavefunctions. We employ only the dominant $\gamma_0 
\gamma_0$ Hamiltonian, without including effects from the 
transverse gluon exchange $\alpha \cdot \alpha$, but the 
comparison with the charmonium states is fair.
We obtain the $s$-wave lepton widths \cite{Estrada:2004dx}
$\Gamma^{nS}_{e^-e^+}\ (keV)$ 
4.4, 2.3, 1.9, 0.98, that show that the wavefunction is being 
artificially forced to shorter distances to accommodate the new 
state, and the widths of the higher radial excitations are larger than 
data by about a factor 2. 

\begin{figure}
\psfig{figure=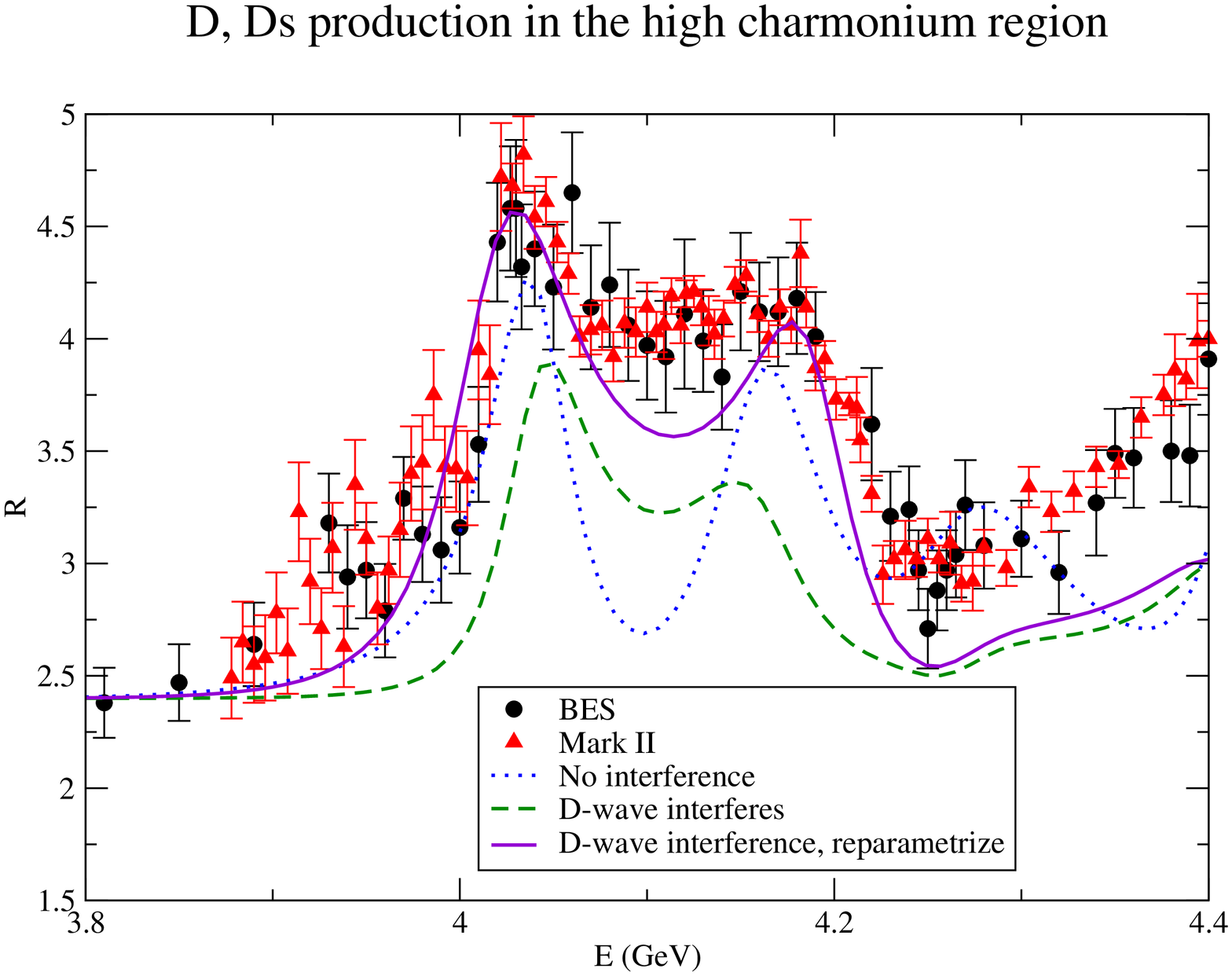,width=2.9in,angle=-90}
\caption{\label{charm1} 
Effect of a new charmonium state on $D^+ D^-$, $D^+_s D^-_s$ 
production at 4260 $MeV$, with parameters from BaBar observation. 
We employ standard PDG parameters for the known resonances (dotted line).
For comparison, the total $R(e^-e^+\to {\rm hadrons})$ is 
given. The non resonant $D$, $D_s$ production as well as light quark 
production (at the 2.4 level) are also added to the calculation for a 
more meaningful comparison.
Observe the spectrum looks qualitatively much better if the $d$-wave 
resonances $\psi(4160)$ and $\psi(4440)$ are taken to have negative
coupling to the double pseudoscalar channel, as calculated by Godfrey and 
Isgur (dashed line).
Finally, reparameterizing the resonances with $M_{\psi(3s)}=4.02 \ GeV$,
$\Gamma_{\psi(3s)}=80 \ MeV$, $\Gamma^{e^-e^+}_{\psi(3s)}=1.5 \ keV$,
$M_{\psi(2d)}=4.19 \ GeV$, $\Gamma^{e^-e^+}_{\psi(2d)}=1.4 \ keV$ (full 
line) suggests leptonic width larger than usually quoted, and in better 
agreement with model calculations.
}
\end{figure}

In figure \ref{charm1} we report a calculation of $D^\pm$, $D^\pm_s$ 
production in $e^-e^+$ scattering. We include a QCD background to account 
for the light quark production, and $D$, $D_s$ meson form factors that 
include the effect of the various $\psi$ resonances. We also recall
\cite{Godfrey:1985xj} that the couplings of the $s$ and $d$-wave 
$q\bar{q}$ $1^{--}$ mesons decaying to a pseudoscalar pair have different 
signs. This interference explains why the $Y(4260)$ has 
not been seen in past searches, as the $s-d$ interference washes it out
of the $DD$, $D_sD_s$ spectrum and it must decay partly into other, less 
favorable channels.

\section{How to experimentally rule out exotic assignments}

We should also consider the possibility of assigning the newly found 
$Y(4260)$ to an exotic multiplet. The new charmonium and charmed 
spectroscopy has yielded several candidates for cryptoexotic tetraquarks,
hidden exotics that have conventional quantum numbers but whose properties
might reveal a large sea quark component. Recent arguments to classify
the various candidates can be found in the literature,
for the $X(3872)$ \cite{Swanson:2004pp}, $D_{SJ}(2632)$ 
\cite{Barnes:2004ay}, $D_S(2320)$ \cite{Szczepaniak:2003vy} and
\cite{Bicudo:2003kj}, or $D_S(2308),(2317)$ \cite{Dmitrasinovic:2005gc}.
An unlikely possibility for $Y(4260)$ is a $J/\psi f_J$ 
loosely bound molecule (assuming isospin 0). A small admixture thereof on 
the other hand explains neatly the observed $J/\psi \pi \pi$ decay.
This can be excluded by increased statistics for the $\pi\pi$ spectrum.
Its proximity to the threshold for $DD^{+}_{0,1,2}$ mesons and other 
channels shows that a sizeable tetraquark or meson molecule should be 
taken into account in this state's wavefunction. This seems to be one of 
the recurrent features of newly found charmed and charmonium states, and 
its understanding is the most pressing task for theorists.

Also of current interest are states that could contain hidden glue and  
now we turn to them. 
The hybrid meson threshold from lattice gauge theory is at around $3.8\ 
 GeV$ \cite{Liao:2002rj},  from many body theory at around $4.3\ GeV$ 
\cite{Estrada:2000hj}. 
The found $Y(4260)$ is close to this second
threshold, and the possibility of forming various $L-S$ combinations in 
a three-body system give four closely spaced  states with various decay 
patterns (see figure (2) in \cite{Estrada:2000hj}). 
Still, this assignment presents two problems.
To form the quantum numbers $1^{--}$, a $p$-wave is needed
either in the $c\bar{c}$ pair (three states) or between this and the 
hidden gluon (one state).
We would then have expected the observed decay channel to be suppressed, 
against a $\chi_c$ $p$-wave meson, as indicated also by the flux tube 
model  
This assignment can be ruled out if the (dominant) open charm modes
do not contain a $p$ wave meson $D^+$.
Finally, a fourth low-lying $c\bar{c}g$ state
has the $p$-wave between the gluon and the quark-antiquark pair. This 
state could then have appreciable branching in the observed final-state 
channel, but the $J/\psi$ would pick up a unit of orbital angular 
momentum respect to the  
pion pair, that should also be in a relative $p$-wave to account for
parity conservation.

Furthermore, let us observe that in the many-body
language, the color octet gluon forces the quark-antiquark pair to be
in a color octet too. The repulsive short range interaction then forces
them apart and makes the formation of a $J/\psi$ in the final
state unlikely. When decaying to hidden charm states,
Hybrid mesons would prefer on these general grounds  a $p$-wave charmonium.
This is consistent with studies of hybrid mesons in the flux tube model 
and employing lattice adiabatic potentials \cite{waidelich}.

The $Y(4260)$ is also massive enough to be a vector glueball. According to
lattice studies \cite{Morningstar:1999rf}, vector glueballs appear in 
the spectrum at $3850(200)\ MeV$. This is in agreement with many-body 
theory calculations \cite{SLAC-PUB-11312}, where the 
minimum Fock space assignment for this oddball would be a three-gluon 
state. This option would require
a tiny reported leptonic width (hence a large branching 
fraction in the $J/\psi\pi\pi$ channel)
 Since the $ggg$ content of the state would be 
flavor blind, an oddball would have a large branching ratio to $\phi \pi 
\pi$. This should be easy to rule out (establish) in the all-charged mode 
$K^+ K^- \pi^+ \pi^-$ with an analysis of the already available data.

\section{Conclusions}

Pending confirmation of this state, we endorse the $Y(4260)$ as the
$\psi(4260)$, corresponding to the low member of the pair $4S-3D$ 
vector charmonium. 
To clarify the assignment, we propose that further studies of this state
with higher statistics attempt to \\
1) Discern whether the $\pi\pi$ subsystem is in an $s$-wave relative to 
the $J/\psi$ instead of a $p$-wave, thus making the $c\bar{c}g$ hybrid 
assignment unlikely. \\
2) Discard the $\phi \pi \pi$ channel typical of  flavor-blind  oddball 
(three-gluon glueball) decay. \\
3) Search the $J/\psi\pi^{\pm}\pi^0$, $J/\psi\pi^0\pi^0$ channel to 
determine the isospin. $I\not =0$ would rule out the simple charmonium 
assignment and make this state a likely tetraquark candidate.  \\
4) Increased statistics should allow the identification of 
the $\psi(4040)$ in the $J/\psi\pi\pi$ spectrum in the same 
experiment, as it corresponds to the (largely) 3S excitation. \\
5) Finally, an attempt to identify $D^{+}_{0,1,2}$ mesons 
is important to identify the role of the nearby $DD^+$ threshold.

The absence of a $J/\psi KK$ signal at the present statistics is not a 
conflict with the  4S assignment, the lepton width and mass, although low, 
are not in blatant disagreement with well established physics.
We finally explain why this resonance has been missed in 
past searches, being in a readily accessible channel with $1^{--}$ 
quantum  numbers, invoking $s$ and $d$ wave interference in the $D$, $D_s$ 
form factors.

\section{Comment on other approaches}

Since the first appearance of this preprint, other authors have exposed 
their views on this state, mostly suggesting exotic assignments that we now 
briefly comment.

In ref. \cite{Liu:2005ay} it has been suggested that a $\chi\rho$
interpretation is likely. We do not see another merit to this conjecture 
than the proximity of the relevant threshold, and the Resonating Group 
Method predicts no special attraction in this channel 
\cite{Ribeiro:1978gx} that would suggest a bound state.

The author of ref. \cite{Zhu:2005hp} assigns the state to a hybrid 
multiplet by discarding other possibilities, in 
particular the charmonium assignment that we here point out is not 
unlikely.
Further, the authors of \cite{Kou:2005gt} support the hybrid assignment. 
They make use of the fourth hybrid state mentioned above where the p-wave
is assigned to the gluon. As explained, this assignment is 
challenged when trying to explain either the $J/\psi\pi\pi$ branching 
fraction or the coupling to the photon probe, since the $c\bar{c}$ pair
is spatially separated. The authors consider the decay of the state into 
$J/\psi \pi\pi$ by the emission of a second physical gluon. This faces 
another conceptual difficulty, as the intermediate channel for the decay is a 
glueball-$J/\psi$ state, at 4.8 GeV by most estimates, which is somewhat 
off-shell and adds to the suppression on the basis of wavefunction 
overlap. The selection rule that the authors prove is aimed at suppressing
$D\bar{D}$ decay and hence making this state narrow. We just observe that 
the $s-d$ wave interference mechanism in conventional charmonium achieves 
a similar effect.

\begin{figure}
\psfig{figure=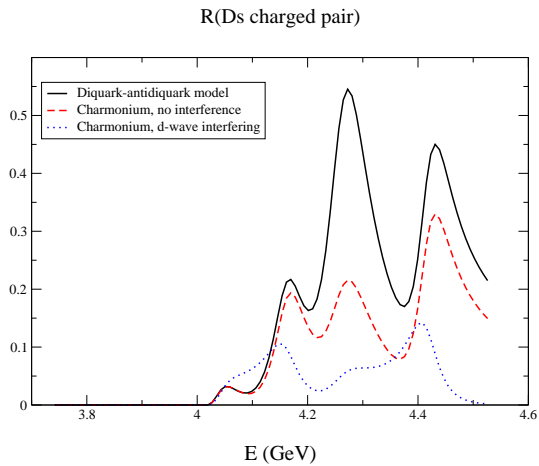,width=2.6in,angle=-90}
\caption{\label{Dsspectrum}
Comparison of tetraquark and charmonium (with and without $d-$wave 
interference) $\frac{\sigma(e^- e^+ \to D_s^+ D_s^-)}{\sigma(e^- e^+ \to 
\mu^+ \mu^-)}$ }
\end{figure}

Finally, another article \cite{Maiani:2005pe} proposes a tetraquark 
$(cs)(\bar{c}\bar{s})$ assignment. Tetraquarks suffer from the well-known 
problem of {\it state inflation}, since there are multiple spin, flavor, 
color and spatial wavefunctions to combine. It is not difficult to find 
suitable candidates for about any meson. A global analysis is 
necessary to discriminate which states do appear in the spectrum. 
We do not find compelling the claim about an $f_0$ state in the $\pi\pi$ 
spectrum since this peaks at high available momenta for other processes, 
it is not visible with the present data, and could point out to a small 
$J/\psi f_0$ admixture and not a tetraquark. More compelling is their 
prediction $\Gamma_{D_s D_s} >> \Gamma_{DD}$, that is however difficult to 
establish experimentally. For comparison, we plot in figure 
\ref{Dsspectrum} the $Ds$ spectrum obtained in a conventional charmonium 
(with and without interference) and tetraquark models. The latter faces 
difficulties similar to the 
other models in terms of explaining a large $J/\psi\pi\pi$ width 
or coupling to a photon probe (since a $p$-wave separates the diquark and 
the antidiquark). 

Therefore we do not see any compeling reason to adopt one of these 
cryptoexotic model assignments since, on a priori reasons, they do not 
solve the conceptual problems that this state causes. The last model has
at least the merit of making a prediction that might be tested.

{\it The author thanks a Fundacion del Amo-Univ.
Complutense fellowship, the hospitality of the SLAC theory
group, and useful conversations with H. Quinn, A. Snyder, 
K. Yi, U. Mallik, G. Hou and M. Herrero. This work is based on a 
long, fruitful and enjoyable collaboration with S. R. Cotanch.
Work partly supported by Spanish MCYT grant FPA 2004-02602,
U. S. DOE Grant DE-FG02-97ER41048.

\end{document}